# The effect of flares on total solar irradiance

Matthieu Kretzschmar[1*], Thierry Dudok de Wit[1], Werner Schmutz[2], Sabri Mekaoui[3], Jean-François Hochedez[4], Steven Dewitte[3]

*[1]LPC2E - Laboratoire de Physique et Chimie de l'Environnement et de l'Espace, UMR 6115 CNRS et Université d'Orléans, 3A Av. de la Recherche Scientifique, Orléans Cedex 2, France.*

*[2] Physikalisch-Meteorologisches Observatorium Davos and World Radiation Center (PMOD/WRC), Dorfstrasse 33, 7260 Davos Dorf, Switzerland.*

*[3] Royal Meteorological Institute of Belgium, Ringlaan 3, 1180 Brussels, Belgium.*

*[4] Solar Influences Data Analysis Centre / Royal Observatory of Belgium, Circular Avenue 3, B-1180 Brussels, Belgium.*

* Corresponding author

*Flares are powerful energy releases occurring in stellar atmospheres [1]. Solar flares, the most intense energy bursts in the solar system, are however hardly noticeable in the total solar luminosity [2,3]. Consequently, the total amount of energy they radiate 1) remains largely unknown and 2) has been overlooked as a potential contributor to variations in the Total Solar Irradiance (TSI)[4], i.e. the total solar flux received at Earth. Here, we report on the detection of the flare signal in the TSI even for moderate flares. We find that the total energy radiated by flares exceeds the soft X-ray emission by two orders of magnitude, with an important contribution in the visible domain. These results have implications for the physics of flares and the variability of our star.*

Flares are sudden bursts of electromagnetic radiation occurring in stellar atmospheres. They are preferentially observed in cool dMe stars where contrast is higher. However, flares have also been observed in other types of stars[1]. In the case of the Sun, imaging capabilities have made it possible to observe flares in the visible domain[5,6] and at short wavelengths, in the X-ray or Extreme UltraViolet (EUV) domains, where the contrast is highest and high quality space instrumentation has been purpose built. However the *total* energy radiated by flares and its spectral distribution remains largely unknown. In fact, the flare signal in the visible domain or in spectrally integrated observations is hindered by the background fluctuations that are caused both by internal acoustic waves and by solar granulation (see fig.1). This explains why among the more than 20,000 flares that occurred in the last solar cycle, only four exceptionally large ones have been identified up to now[7] in the TSI, i.e. the flux of solar light received at all wavelengths at the top of the Earth's atmosphere. The TSI is the primary energy input to Earth, so identifying the impact of flares on TSI is of fundamental importance in several aspects: it provides an important constraint for the physics of flares, it makes it possible to compare flares on other stars and it offers an estimate of flare contribution to solar luminosity.

To identify the effect of flares on the TSI, the aforementioned fluctuations must be somehow cancelled out; to do so, we performed a conditional average analysis[8], which is also known as superposed epoch analysis. The procedure involves 1) selecting a set of flares, 2) extracting an excerpt of the irradiance time series around the flare peak time for these events and 3) averaging (or superposing) these smaller time series in such a way that flares occur at the same time in all time series.

We applied this procedure to measurements of the TSI recorded by the PMO and DIARAD radiometers of the VIRGO[9] experiment onboard the SOHO mission between 1996 and 2007, i.e. during the whole 23$^{rd}$ solar cycle. We used the Soft X-Ray (SXR) flux observed by the GOES satellites in the spectral range 0.1-0.8 nm to retrieve the flare information and in particular the peak time. The flare amplitude is given by its X-ray class: the letter indicates a power of ten ('X' is -4, 'M' is -5, 'C' is -6, etc.). A flare of class M4.5 thus corresponds to a SXR peak flux of $I^{SXR}$=4.5 $10^{-5}$ W/m$^2$. We considered only flares occurring at heliocentric angle less than 60° (to limit line-of-sight effects) and of class greater than M3 in order to stay sufficiently above the instrumental background. To maximize the flare signal, we ranked the flares from the largest to the smallest using the SXR flux peak $F^{SXR}$ measured by the GOES satellites: $F_1^{SXR} > F_2^{SXR} > ... > F_i^{SXR} > ... > F_N^{SXR}$, and we applied the analysis to several sets of flares that are characterized by two values: the largest flare of the set (noted by its rank $k$) and the total number $n$ of flares inside the set (the smallest flare of the set then being of rank $k + n - 1$). For each flare, we selected a TSI interval of $\tau$=7 hours in such a way that the flare peak time is situated 180 minutes after the interval begins. These short intervals were divided by their time average before being superposed, so that the average relative TSI variations resulting from the analysis can be expressed as $\bar{I}(k,n,t) = 10^6 \cdot \frac{1}{n} \sum_{i=k}^{k+n-1} \left( \frac{I_i(t)}{\langle I_i(t) \rangle_\tau} - 1 \right)$, where $I_i(t)$ is the measured TSI time series during the flare of rank $i$ and $\bar{I}(k,n,t)$ is expressed in parts per million (ppm). The analysis is based on the fact that although the TSI value at peak time $I_i(t_{peak})$ cannot be distinguished from the value at other times, it includes a small contribution $I_i^f$ due to the flare. At times where no flare emission occurs (far from $t_{peak}$), the incoherent fluctuations are averaged out and the value $\bar{I}(t)$ goes to zero as $\sigma / \sqrt{n}$, where $\sigma$ is the standard deviation of the original time series. At the peak of the flare, the coherent contribution of the flares causes the average value to differ from zero and to tend towards the average relative TSI increase $\bar{I}(k,n,t_{peak}) = \overline{I^f}(k,n) = \frac{1}{n} \sum_{i=k}^{k+n-1} I_i^f$. This supposes that other phenomena such as acoustic waves and active region evolution are not in phase with the flares and therefore do not contribute to the flare signature.

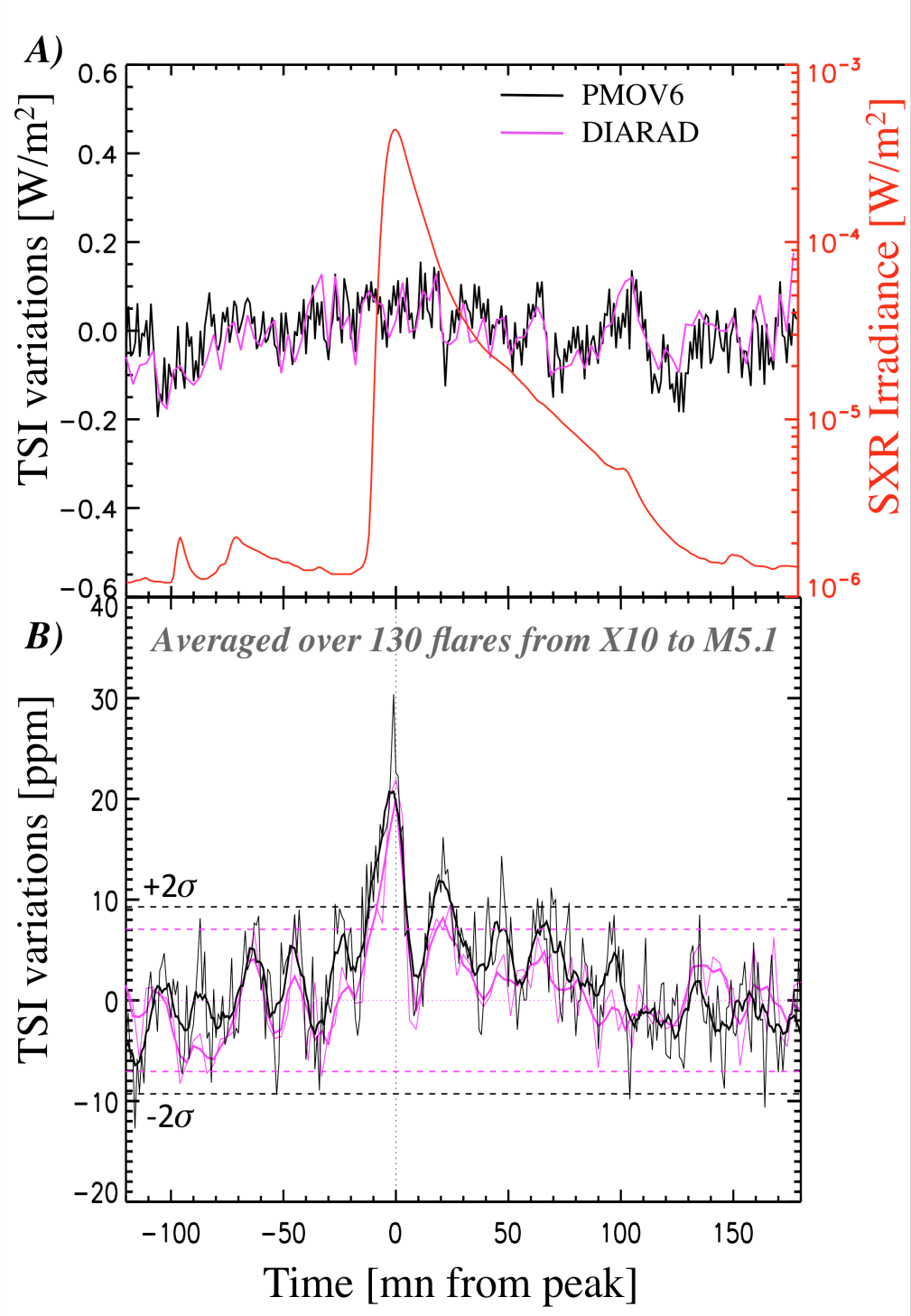

**Fig. 1: Detection of flares in the total solar irradiance through superposed epoch analysis.** ***Panel A: Variation of total solar irradiance during a single X4 flare***, which occurred on 26th November 2000. The black and pink curves correspond respectively to the TSI measured by the PMOV6 and the DIARAD radiometers. The red curve represents the SXR irradiance as measured by the GOES spacecraft. The flares show large contrasts in the EUV and SXR domain but are barely perceptible - if at all - in the TSI. The standard deviation $\sigma$ of the PMOV6 time series is 0.07 W/m$^2$, which corresponds to about 50 ppm. ***Panel B:*** **Averaged TSI variations** in ppm for 130 of the largest solar flares of the last solar cycle occurring at $\theta$ <60°, obtained through a superposed epoch analysis. Colour-coding is the same as in panel A. The time step is respectively 3 minutes and 1 minute for DIARAD and PMOV6. Thick lines are obtained by averaging over 6-minutes. The dashed lines correspond to the 95% confidence level (twice the standard deviation computed everywhere except for 1 hour around the flare). The averaging reduces the fluctuation level and reveals the coherent contribution of flares. ***Panel C:*** **Effect of the flares position**. In black, the TSI average for the 130 largest solar flares of the last solar cycle occurring at $\theta$ <60° (same as in panel B), and in blue, the TSI average for the 130 largest solar flares occurring at all $\theta$.

We first considered the conditional averaging of 130 very large flares of the last solar cycle, shown in panel B of fig. 1. Firstly it should be noted that the fluctuation level is reduced from its initial level of about $\sigma$ = 50 ppm to about $\sigma(n) = 50/\sqrt{n}$ ppm in the averaged curve. More importantly, the average TSI clearly reveals a statistically significant peak at the time of the flare. Most of the TSI increase happens before the key time, which shows that this emission comes mainly from the impulsive phase.

This remarkable result remains valid when we consider flares of lower amplitude (see fig. 2), for the first time revealing a significant impact on the TSI, not only for the few largest flares but also for the numerous smaller ones. Fig. 2 shows four exclusive sets of flares with a clear signature in the TSI at the time of the X-ray flare. For the smallest flare in panel D, the signal is less clear in the DIARAD data, which can at least in part be explained by the smaller duty cycle of the instruments. It is important to note that the apparent periodic patterns in fig.2 and fig.1 are due to the random superposition of the pseudo-periodic TSI oscillations at the typical frequency of p-modes (3mHz); these patterns are more prominent in the smoothed curves but remain below the $2\sigma$ level, and can thus hardly be interpreted as actual signatures of physical processes related to flares. Figures 1 and 2 show the first important finding of this study, namely that the total radiative output of the Sun is sensitive to both large and small flares.

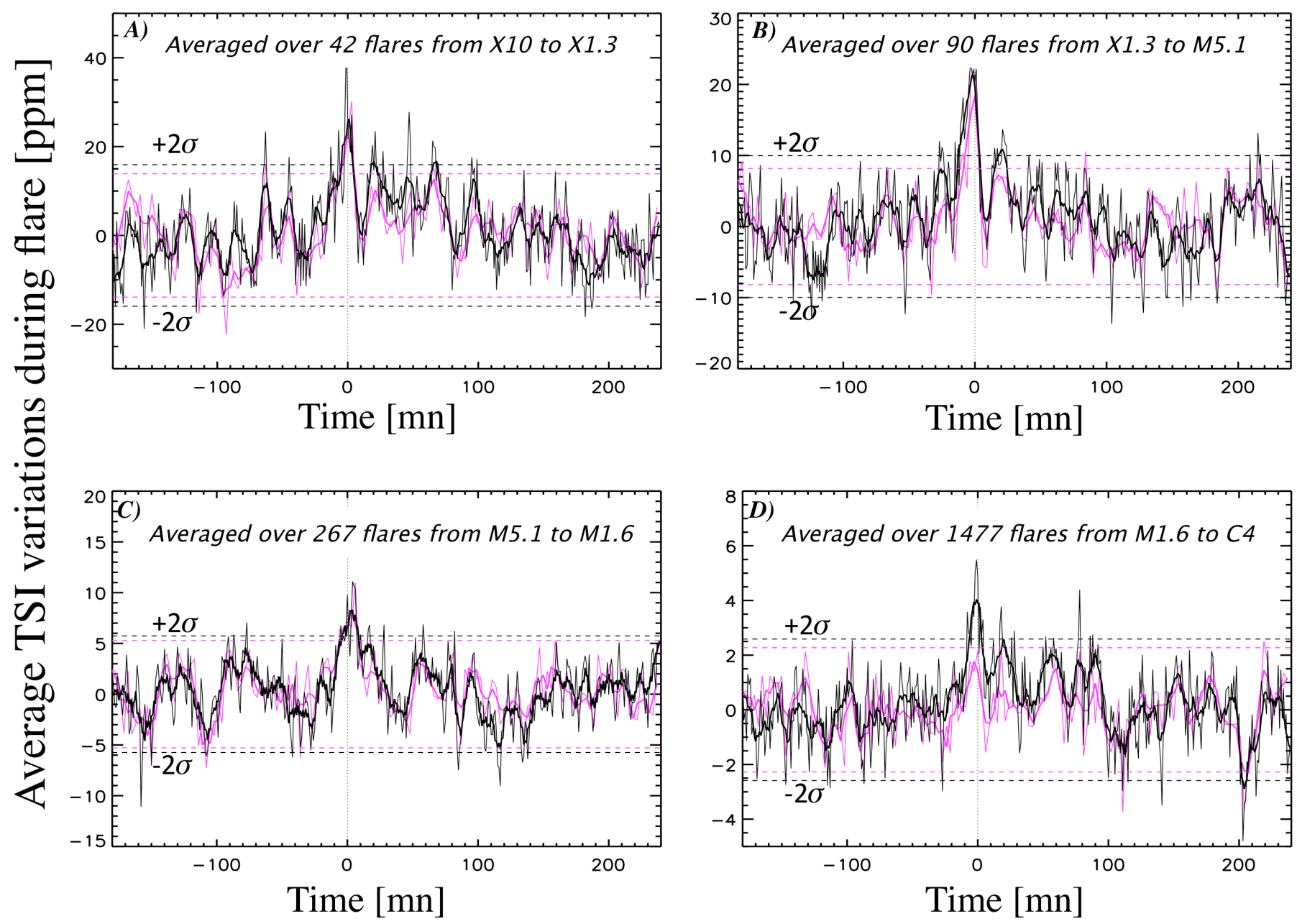

**Fig. 2: Averaged TSI variations during flares of smaller amplitudes.** Panels A to D show the TSI time series averages over four exclusive sets of solar flares of decreasing amplitude. The black and pink curves correspond respectively to the TSI measured by the PMOV6 and the DIARAD radiometers. The thick lines are 6 minutes running averages. The dashed lines correspond to the 95% confidence level for the thin curves (twice the standard deviation computed everywhere except for 1 hour around the flare).

Finally, let us look at another quantity of great interest: namely the total energy radiated by flares. This energy was simply estimated by integrating the averaged TSI variations over 20 minutes around the peak time and multiplying it by the factor $1365.10^7*(1AU)^2*1.4\pi$, the factor 1.4 being the best compromise between a uniform angular distribution (factor $2\pi$) expected for optically thin wavelengths and a Lambertian distribution (factor $\pi$) [7]. The fixed integration time avoids errors caused by low signal-to-noise ratio in determining the flare duration.

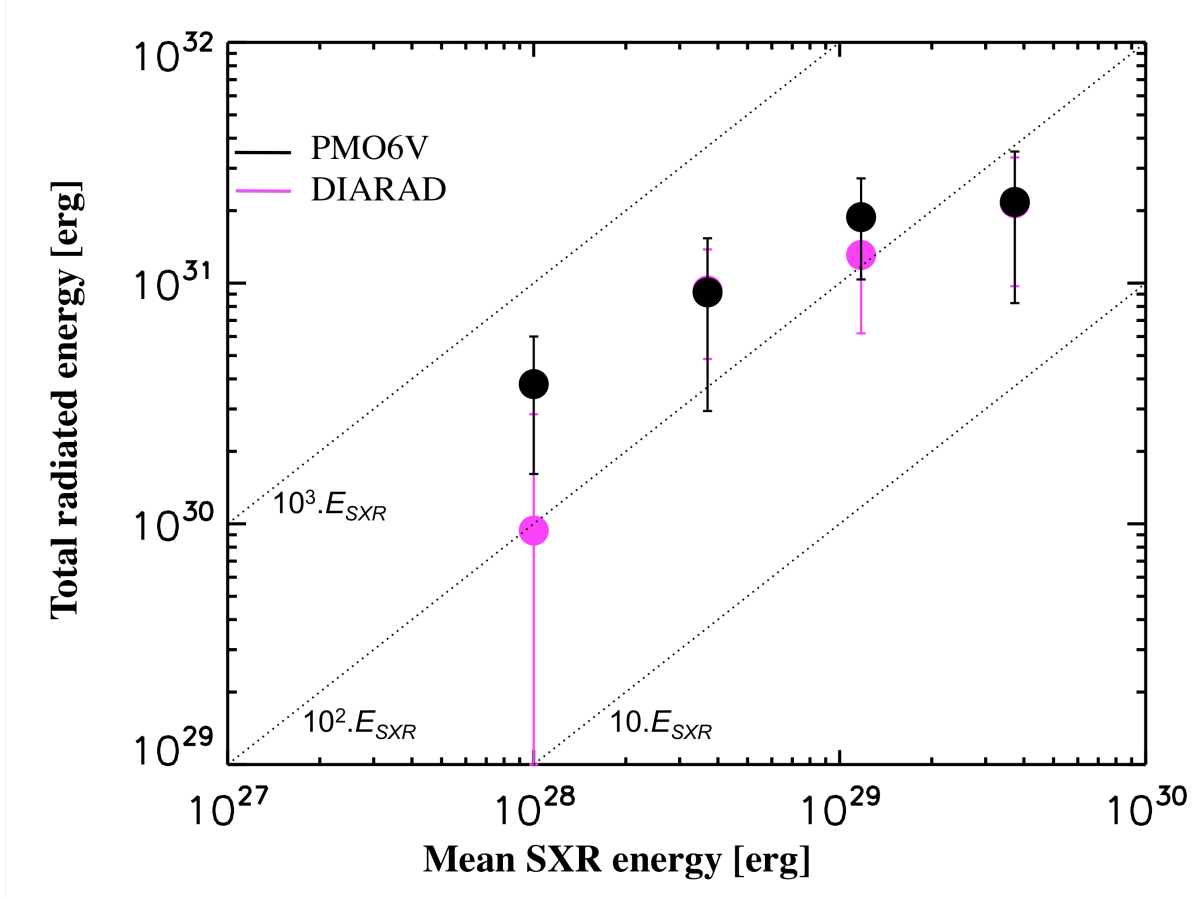

**Fig. 3: Total energy radiated by solar flares versus the energy radiated in X-rays**. The black and pink symbols correspond respectively to estimates using PMO6V and DIARAD. Each point corresponds to a panel of fig.2. The error bars correspond to the $1\sigma$ dispersion in the averaged time series and are only indicative; the uncertainties are indeed larger since part of the flare profile is below the noise level.

The main conclusion that can be drawn from the results, which are shown in fig.3, is that the total energy radiated by solar flares exceeds by two orders of magnitude the energy radiated in the SXR domain; this scaling holds for two decades of flare amplitudes. This confirms the recent surprisingly large estimate for the largest flares only[3,7], while considerably extending its validity range; this result affects the whole energy budget of flares as well as the relation between flares and TSI. It should also be noted that the values shown in fig. 3 could actually be underestimated since flares usually emit over a longer time than that used for the integration. The relatively small contribution of X-rays to the total emission points towards a dominant energy contribution of longer wavelengths, mainly in the near UV and in the visible domain. This is supported by extending our analysis to spectral irradiance measurements carried out in the visible domain by the SPM sensors of the VIRGO instrument, which shows that flares produce an increase in these visible channels that is similar in time and amplitude to the one observed in the TSI (not shown). It thus appears that the contribution of the visible and near UV emission to the total flare emission are large enough to make the signature of flares detectable in the total radiative output of our star.

1. Haisch, B., Strong, K.T. & Rodono, M. Flares on the sun and other stars. *Annual Review of Astronomy and Astrophysics* **29**, 257-324 (1991)
2. Hudson, H. S. & Willson, R. C. Upper limits on the total radiant energy of solar flares. *Solar Physics* **86**, 123-130 (1983)
3. Woods, T. N., Eparvier, F.G., Fontenla, J., *et al.* Solar irradiance variability during the October 2003 solar storm period. *Geophysical Research Letter* **31**, 10802 (2004)
4. Fröhlich, C. & Lean, J. Solar radiative output and its variability: evidence and mechanisms. *The Astronomy and Astrophysics Review*. **12**, 273-320 (2004)
5. Carrington, R.C. Description of a Singular Appearance seen in the Sun on September 1. *Monthly Notices of the Royal Astronomical Society* **20**, 13-15 (1859).
6. Neidig, D. F. The importance of solar white-light flares *Solar Physics* **121**, 261-269 (1989),
7. Woods, T. N., Kopp, G., & Chamberlin, P. C. Contributions of the solar ultraviolet irradiance to the total solar irradiance during large flares. *J. Geophys. Res.* **111**, A10S14 (2006)
8. Johnsen, H., Pécseli, H. L. & Trulsen, J. Conditional eddies in plasma turbulence. *Physics of Fluids* **30** (7), 2239-2254 (1987).
9. Frohlich, C., Andersen, B. N., Appourchaux, T. et al., First Results from VIRGO, the Experiment for Helioseismology and Solar Irradiance Monitoring on SOHO. *Solar Physics* **170**, 1-25 (1997)

**Acknowledgments** This work has received funding from the European Community's Seventh Framework Programme (FP7/2007-2013) under the grant agreement n° 218816 (SOTERIA project, www.soteria-space.eu). J.F.H., T. DdW. and M. K. also acknowledge financial support from the French-Belgium partnership program “Tournesol”.

**Author Contribution** S. D., S. M., W. S. and J.F.H. were involved in the design of the study. T. DdW. was involved in the analysis of the data. M. K. analysed the data and drafted the manuscript. All authors discussed the results and commented on the manuscript.